\documentclass[aps,prc,twocolumn,superscriptaddress,showpacs]{revtex4-1}
\usepackage{graphicx}
\usepackage{hyperref}
\usepackage{amssymb}
\usepackage{color}
\usepackage[normalem]{ulem}
\usepackage{lineno}

\begin{document}

%\preprint{}

\title{Elliptic and Triangular Flow of Identified Particles from the AMPT Model at RHIC Energies}



\author{Xu Sun}
%\email[]{xsun@hit.edu.cn}
\affiliation{Department of Physics, Harbin Institute of Technology, Harbin 150001, China}

\author{Jianli Liu}
%\email[]{liujianlihit@gmail.com}
\affiliation{Department of Physics, Harbin Institute of Technology, Harbin 150001, China}

\author{Alexander Schmah}
%\email[]{aschmah@lbl.gov}
\affiliation{Nuclear Science Division, Lawrence Berkeley National Laboratory, Berkeley, California 94720, USA}

\author{Shusu Shi}
\affiliation{Key Laboratory of Quarks and Lepton Physics (MOE) and Institute of Particle Physics, Central China Normal University, Wuhan, 430079, China}

\author{Jingbo Zhang}
%\email[]{jinux@hit.edu.cn}
\affiliation{Department of Physics, Harbin Institute of Technology, Harbin 150001, China}

\author{Hanzhi Jiang}
\affiliation{School of Astronautics, Harbin Institute of Technology, Harbin 150001, China}

\author{Lei Huo}
\email[]{lhuo@hit.edu.cn}
\affiliation{Department of Physics, Harbin Institute of Technology, Harbin 150001, China}


\date{\today}

\setlength\linenumbersep{0.10cm}

\begin{abstract}
 The elliptic flow ($v_{2}$) at $\sqrt{s_{\rm NN}} = $ 11.5, 39, and 200 GeV and triangular flow ($v_{3}$) at $\sqrt{s_{\rm NN}} = $ 200 GeV of identified particles ($\pi^{\pm}, K^{\pm}, K^{0}_{S}, p, \bar{p}, \phi, \Lambda$ and $\bar{\Lambda}$) from 0\%--80\% central Au+Au collisions are analyzed using a multiphase transport (AMPT) model. It is shown that the experimental results from the $\eta$-sub event plane method can be reproduced with a parton scattering cross section between 1.5 and 3 mb. We also studied the differential and integrated $v_{2}/v_{3}$ ratios and conclude that they are anti-correlated with the parton scattering cross section.
%  And no mass (particle type) dependence is observed for integrated $v_{2}/v_{3}$ ratio.
%  The possible relation to viscous effect is discussed.
\end{abstract}

\pacs{25.75.-q, 25.75.Ld}
%\keywords{}

\maketitle

\section{\label{sec:intro} Introduction}
One of the main goals of heavy-ion collision experiments at the Relativistic Heavy Ion Collider (RHIC) is to study the properties and formation of the Quark Gluon Plasma (QGP). The study of the azimuthal anisotropy of emitted particles, based on Fourier decomposition, is considered to be one of the most important tools to investigate the hot and dense matter created in heavy-ion collisions~\cite{Voloshin:2008dg}. Several interesting observations of the second harmonic Fourier coefficient $v_{2}$, also called elliptic flow, have been reported during the past decade by using the data from the top RHIC heavy-ion energy of $\sqrt{s_{\rm NN}} = $ 200 GeV~\cite{Sorensen:2009cz,Adams:2005dq,Adcox:2004mh}. At low transverse momenta ($p_{\rm T} < $ 2.0 GeV/$c$), a mass ordering of $v_{2}(p_{\rm T})$  was observed~\cite{Adler:2001nb,Adams:2003am}, which can be understood within a hydrodynamic framework~\cite{Huovinen:2001cy}.
At intermediate transverse momenta (2.0 $ < p_{\rm T} < $ 6.0 GeV/$c$), a Number-of-Constituent Quark (NCQ) scaling~\cite{Molnar:2003ff} was observed~\cite{Adams:2005zg,Abelev:2007rw,Abelev:2010tr,Abelev:2007qg}.
The NCQ scaling was used to conclude that the relevant degrees of freedom in the created matter are quarks and gluons.
In addition to measurements at top RHIC energies, a Beam Energy Scan (BES) program has been carried out in the years 2010, 2011 and 2014 at to study the structure of the QCD phase diagram. Au+Au collisions were recorded at $\sqrt{s_{NN}} = $ 7.7, 11.5, 14.5, 19.6, 27, 39, and 62.4 GeV. The mass ordering in the low $p_{\rm T}$ region and NCQ scaling at intermediate transverse momenta were also observed BES energies~\cite{Adamczyk:2013gv,Adamczyk:2013gw}.

In the recent years, the third harmonic Fourier coefficient $v_{3}$, called triangular flow, has attracted more attention and was studied intensively~\cite{Alver:2010gr,Teaney:2010vd}. It is thought to be related to the near-side ridge structure observed in two particle correlation measurements~\cite{Alver:2010gr}. The triangular flow seems to be more sensitive to the viscous effects than $v_{2}$~\cite{Alver:2010dn}, but less sensitive to the collision centrality~\cite{Bhalerao:2011yg}. Thus, $v_{3}$ is a natural choice  to study initial fluctuations and viscosity effects.
Similar flow patterns as for $v_{2}$ were recently observed in $v_{3}$ measurements~\cite{Adamczyk:2013waa,Sun:2014yqi}.

Viscous relativistic hydrodynamic calculations~\cite{Retinskaya:2013gca,Retinskaya:2014zea} suggest a combined analysis of elliptic and triangular flow data to put tight constraints on the initial states of nucleus-nucleus collisions.
% if the hydrodynamics response to the initial state is linear.
Other hydrodynamic model calculations~\cite{Lang:2013oba} predict a stronger damping of the higher harmonic flow coefficients (n $>$ 2) relative to $v_{2}$ at high transverse momenta. A constant ratio of $v_{2}(p_{\rm T})/v_{3}(p_{\rm T})$ at high $p_{\rm T}$ is predicted, where high means $p_{\rm T}/m_{\rm T} > \bar{v}_{max}$ and $\bar{v}_{max}$ is the highest flow velocity.
Thus, by combining elliptic and triangular flow, more information about the initial state and the influence of the viscous effect can be collected.

In this paper, the differential and integrated ratio of $v_{2}/v_{3}$ of identified particles ($\pi^{\pm}, K^{\pm}, K^{0}_{S}, p, \bar{p}, \phi, \Lambda$ and $\bar{\Lambda}$) are studied with the AMPT model. This paper is organized as follows. Section~\ref{sec:ep_ampt} gives a brief introduction to the $\eta$-sub event plane method and the AMPT model. In section~\ref{sub:ampt_flow}, the collective flow calculations ($v_{2}$ and $v_{3}$) from the AMPT model are presented.  The differential and integrated $v_{2}/v_{3}$ ratios are discussed in Sec.~\ref{sub:ampt_diff} and Sec.~\ref{sub:ampt_inte}. A summary is given in Sec.~\ref{sec:sum}.

\section{\label{sec:ep_ampt} $\eta$-sub Event Plane Method and the AMPT Model}

\subsection{\label{sub:ep} $\eta$-sub Event Method}
The event plane method is one of the most widely used methods to analyze anisotropic flow in heavy-ion collisions~\cite{Poskanzer:1998yz}. In non-central Au+Au collision, the overlap region has an almond shape in the coordinate space. As the system evolves, the pressure gradient pushes the anisotropy from coordinate space to momentum space, therefore the produced particles have an anisotropic distribution in momentum space. The azimuthal distribution of the produced particles can be written as~\cite{Voloshin:1994mz}:
\begin{equation}
  \label{eq:EP}
  E\frac{d^{3}N}{dp^{3}} = \frac{1}{2\pi}\frac{d^{2}N}{p_{\rm T}dp_{\rm T}dy}(1+\sum_{n=1}^{\infty}2v_{n}^{\rm obs}\cos[n(\phi-\Psi_{n})]),
\end{equation}
where $\phi$ is the azimuthal angle of a particle,  $v_{n}^{\rm obs}$ is the observed n-th harmonic flow., and $\Psi_{n}$ is the n-th harmonic event plane angle reconstructed by the produced particles, defined as:
\begin{equation}
  \label{eq:Psi}
  \Psi_{n} = \frac{1}{n}[\tan^{-1}\frac{\sum_{i}w_{i}\sin(n\phi_{i})}{\sum_{i}w_{i}\cos(n\phi_{i})}].
\end{equation}
The sum goes over all charged particles in the event, $w_{i}$ is a weight applied to optimize the event plane resolution, see~\cite{Poskanzer:1998yz} for details. It is worth to note that the range of n-th harmonic event plane angle is 0 $ \leqslant \Psi_{n} < $ 2$\pi$/n.

In this method the n-th observed harmonic flow can be corrected for statistical effects by the n-th harmonic event plane resolution:
\begin{equation}
  \label{eq:vn}
  v_{n} = v_{n}^{obs}/R_{n},
\end{equation}

with

\begin{equation}
  \label{eq:res}
  R_{n} = \left<\cos[n(\Psi_{n}-\Psi_{nR})]\right>.
\end{equation}
Here $\left<\dots\right>$ denotes the average over all particles in all events and $\Psi_{nR}$ is the n-th real event plane angle~\cite{Xiao:2011ti}.
The real event plane means the participant plane~\cite{Kharzeev:2000ph}, which can not be achieved due to finite multiplicity of an event.

To reduce the short range "non-flow" effects such as HBT correlations, a $\eta$-sub event plane method was introduced~\cite{Adamczyk:2013gw}. The $\eta$-sub event plane method divides the event into two independent sub-events with different pseudo rapidity ($\eta$) ranges. An additional gap of 0.1 in pseudo rapidity is added to further avoid short range correlations. The procedure results in a positive and negative $\eta$-sub event plane.
The n-th observed harmonic flow can now be calculated in respect to the $\eta$-sub event planes. As a consequence, the $\eta$-sub event plane resolution~\cite{Poskanzer:1998yz} is used instead of full event plane resolution:
\begin{equation}
  \label{eq:sub}
  R_{n}^{sub} = \sqrt{\left<\cos[n(\Psi_{n}^{a}-\Psi_{n}^{b})]\right>},
\end{equation}
where the $\Psi_{n}^{a}$ and the $\Psi_{n}^{b}$ are the n-th harmonic event plane angles for the positive and negative $\eta$-sub ranges.

\subsection{\label{sub:ampt} AMPT Model}

AMPT is a transport model which consists of four main components: the initial conditions, partonic interactions, conversion from partonic to hadronic matter, and hadronic interactions~\cite{Lin:2004en}. It has two versions to deal with different scenarios which are default AMPT (AMPT Def in following figures) and string melting AMPT  (AMPT Str in following figures). The initial conditions are generated by the HIJING (Heavy Ion Jet Interaction Generator) model~\cite{Gyulassy:1994ew,Wang:1990qp,Wang:1991hta}. HIJING includes only two body nucleon-nucleon interactions and generates mini jets and excited strings through hard processes and soft processes separately. 

Excited strings are treated differently in the two AMPT versions. In default AMPT, excited strings are combined to hadrons according to the Lund string fragmentation model they further go through a hadronic interaction stage~\cite{Lin:2004en}.
In string melting AMPT, excited strings first convert to partons, i.e. melting, and then go through a partonic interaction stage with original soft partons. The partonic interactions for both default AMPT and string melting AMPT  are described by ZPC (Zhang's Parton Cascade) model~\cite{Zhang:1997ej}. In the final stage of the ZPC model, partons in default AMPT are recombined with parent strings and hadronize via the Lund string fragmentation model.
However, in  string melting AMPT, the hadronization of partons is described by a coalescence model.
After hadronization, the hadronic interactions are modelled by ART (A Relativistic Transport) model~\cite{Li:1991mr,Li:1995pra}.

In previous studies it was found that a large parton scattering cross section (approximately 6 mb or 10 mb) is needed to reproduce the observed elliptic flow of charged hadrons~\cite{Lin:2001zk,Chen:2004dv}. In these studies, the elliptic flow was calculated with respect to the reaction plane (determined by the beam axis and impact parameter direction), which is 0 in the AMPT model, and this might underestimate the elliptic flow observed in the experiment data, which measured relative to the event plane but not the reaction plane~\cite{Poskanzer:1998yz}. 
As shown in Ref.~\cite{Xu:2011fe}, both the charged particle multiplicity and elliptic flow can be reproduced with a smaller parton scattering cross section (1.5 mb) by using the default values for the parameters in the Lund string fragmentation function. 

In this paper, approximately 20 million events for 0\%--80\% central Au+Au collisions at $\sqrt{s_{\rm NN}} = $ 11.5, 39, and 200 GeV with default AMPT  (v1.25) and string melting AMPT (v2.25) were generated. Three different parton scattering cross sections (1.5 mb, 3 mb, and 6 mb) are used in the string melting version of AMPT. The parameters in the Lund string fragmentation function are set to the default values for AMPT  with 1.5 mb parton scattering cross section (the values can be found at Table I in Ref.~\cite{Xu:2011fe}). Furthermore the $\eta$-sub event plane method is used to calculate elliptic and triangular flow.
%-----------------------------------------------------------------------------------------------
% figure 1
\begin{figure*}[t!]
  \centering
  \includegraphics[width=0.8\textwidth]{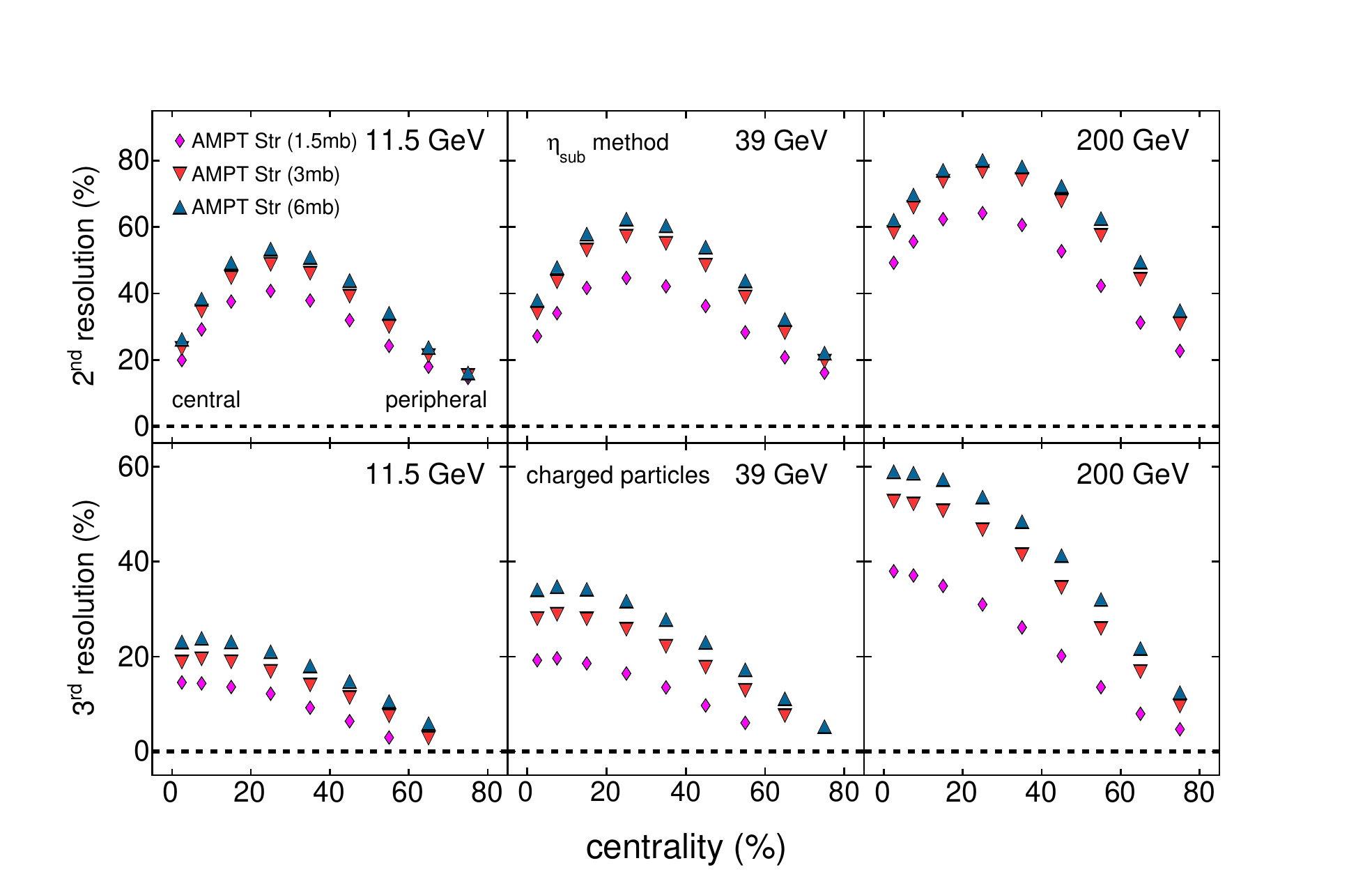}%
  \caption{\label{fig:res}(Color online) Second harmonic (top panel) and third harmonic (bottom panel) event plane resolution from the AMPT string melting model with 1.5 mb (AMPT Str (1.5 mb)), 3 mb (AMPT Str (3 mb)) and with 6 mb (AMPT Str (6 mb)) parton scattering cross section from 0\%-80\% central Au+Au collisions at $\sqrt{s_{\rm NN}} = $ 11.5, 39 and 200 GeV. The analysis method is $\eta$-sub event plane method, and the event plane is reconstructed by charged particles, see text for details.}
\end{figure*}
%-----------------------------------------------------------------------------------------------

\begin{figure*}[ht]
\begin{minipage}[b]{0.45\linewidth}
\centering
\makebox[\linewidth]{%
\includegraphics[width=0.88\textwidth]{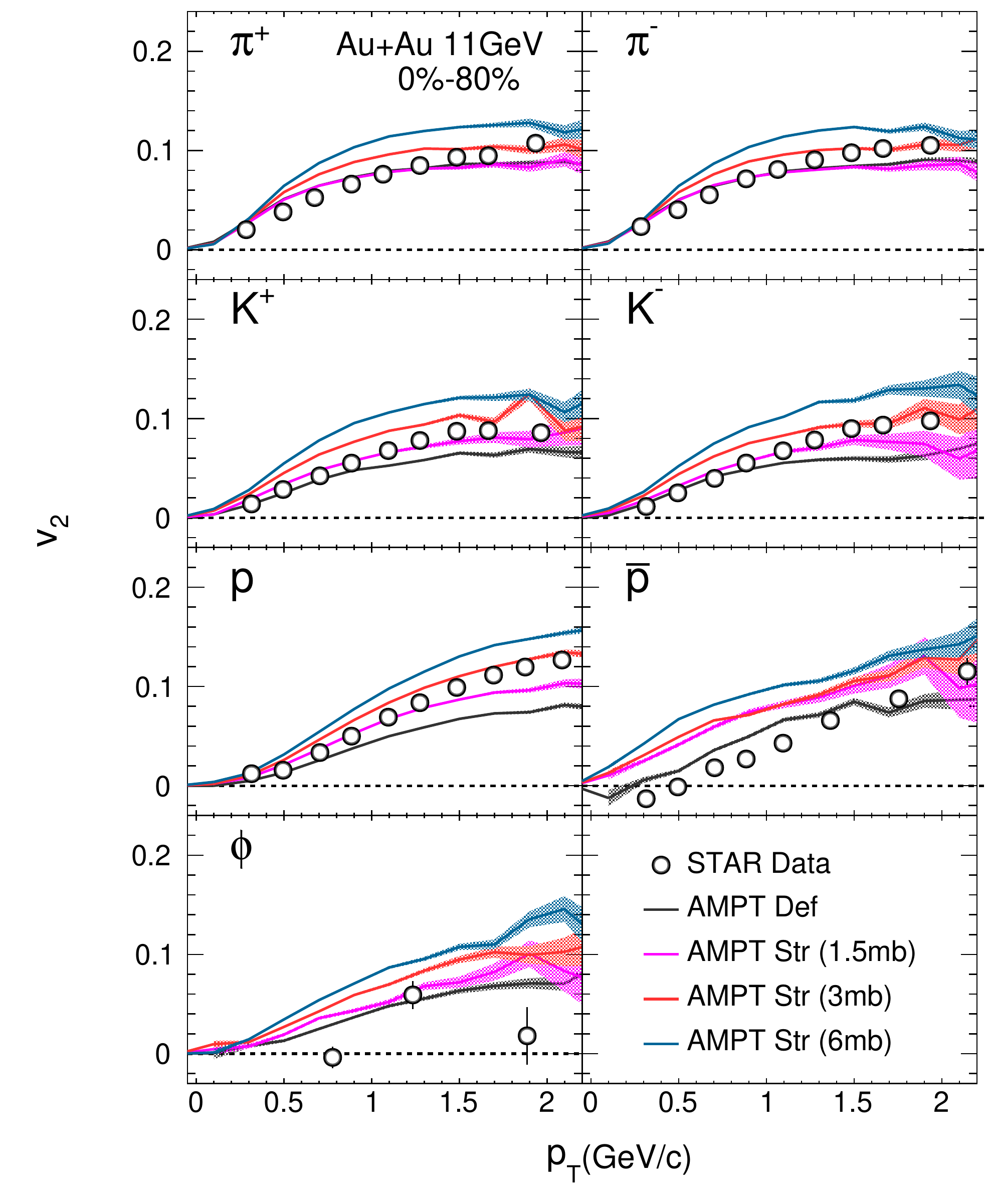}}
 \caption{\label{fig:v2_11} (Color online) $v_{2}(p_{\rm T})$ results at $\sqrt{s_{\rm NN}} = $ 11.5 GeV from 0\%--80\% central Au+Au collisions for identified particles ($\pi^{\pm}, K^{\pm}, p, \bar{p}$ and $\phi$) in comparison to STAR data.}
\end{minipage}
\hspace{0.5cm}
\begin{minipage}[b]{0.45\linewidth}
\centering
\makebox[\linewidth]{%
\includegraphics[width=0.88\textwidth]{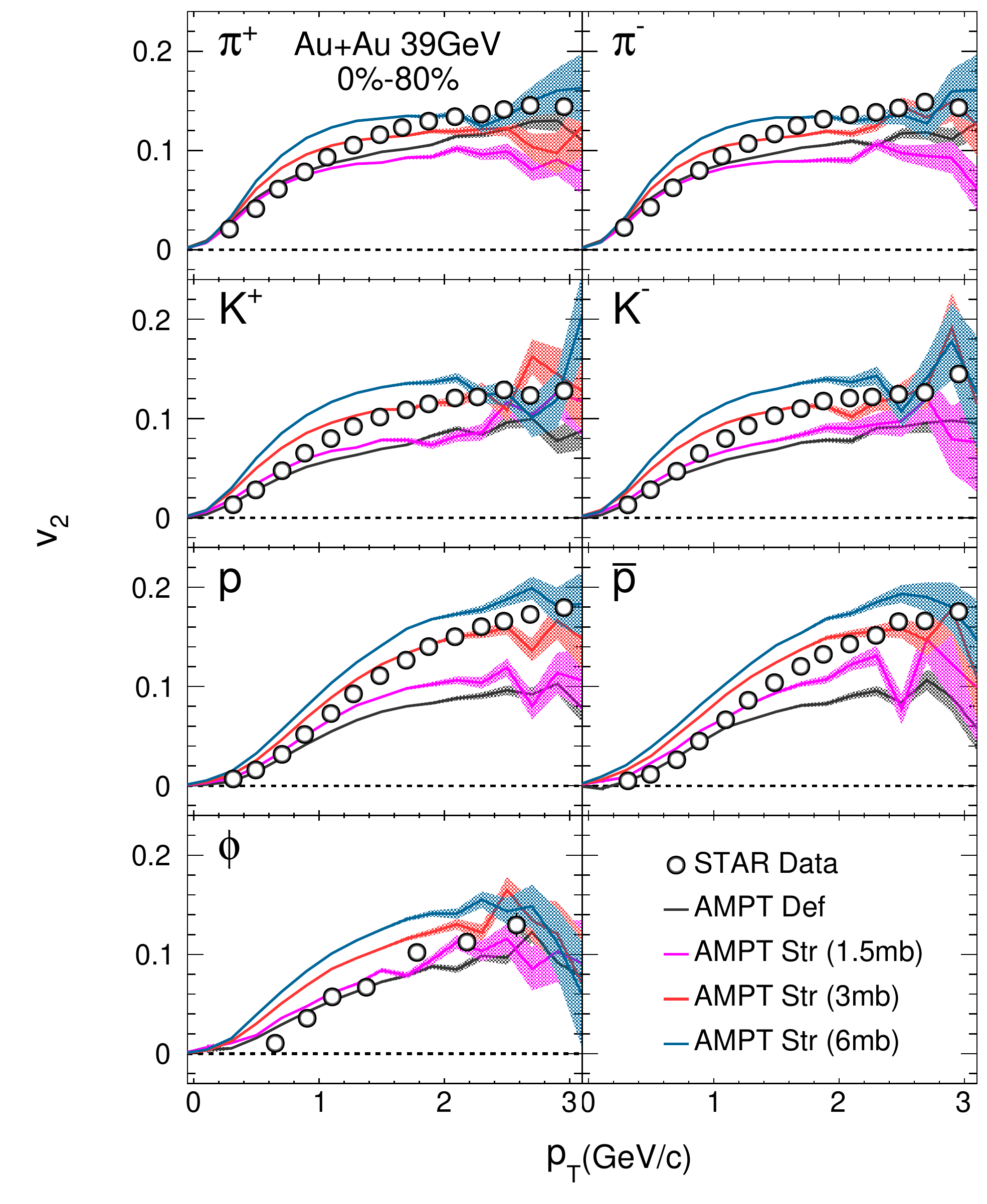}}
  \caption{\label{fig:v2_39} (Color online) $v_{2}(p_{\rm T})$ results at $\sqrt{s_{\rm NN}} = $ 39 GeV from 0\%--80\% central Au+Au collisions for identified particles ($\pi^{\pm}, K^{\pm}, p, \bar{p}$ and $\phi$) in comparison to STAR data.}
\end{minipage}
\end{figure*}

%%-----------------------------------------------------------------------------------------------
%% figure 2
%\begin{figure}[h!]
%  \centering
%  \includegraphics[width=0.4\textwidth]{./fig2_11GeV.pdf}%
%  \caption{\label{fig:v2_11} (Color online) $v_{2}(p_{\rm T})$ results at $\sqrt{s_{\rm NN}} = $ 11.5 GeV from 0\%--80\% central Au+Au collisions for identified particles ($\pi^{\pm}, K^{\pm}, p, \bar{p}$ and $\phi$) compare to STAR data. The open symbols are STAR data and the lines are from AMPT calculation.}
%\end{figure}
%%-----------------------------------------------------------------------------------------------
%%-----------------------------------------------------------------------------------------------
%% figure 3
%\begin{figure}[h!]
%  \centering
%  \includegraphics[width=0.4\textwidth]{./fig3_39GeV.pdf}%
%  \caption{\label{fig:v2_39} (Color online) $v_{2}(p_{\rm T})$ results at $\sqrt{s_{\rm NN}} = $ 39 GeV from 0\%--80\% central Au+Au collisions for identified particles ($\pi^{\pm}, K^{\pm}, p, \bar{p}$ and $\phi$) compare to STAR data. The open symbols are STAR data and the lines are from AMPT calculation.}
%\end{figure}
%%-----------------------------------------------------------------------------------------------
%-----------------------------------------------------------------------------------------------
% figure 4
\begin{figure}[h!]
  \centering
  \includegraphics[width=0.4\textwidth]{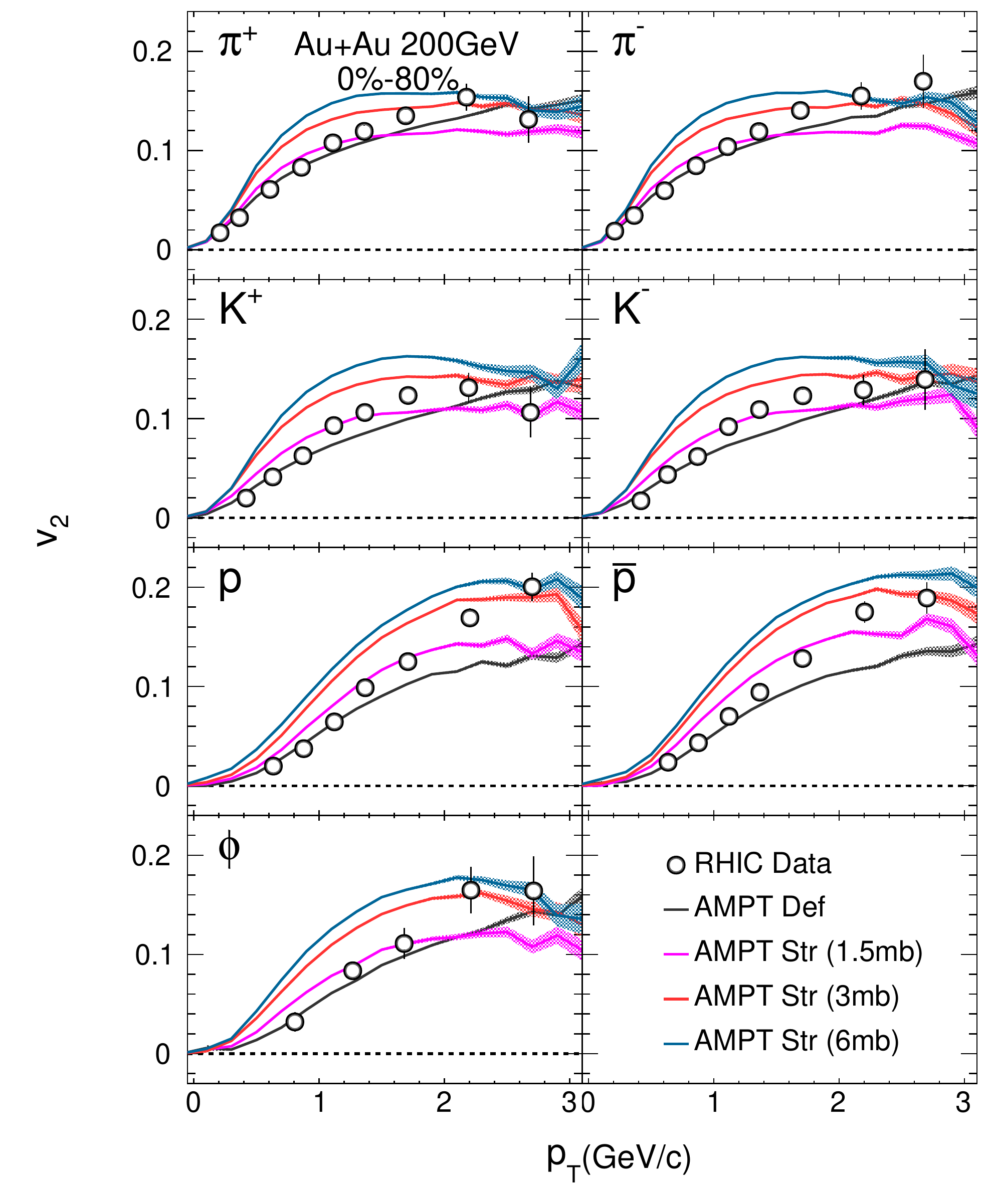}%
  \caption{\label{fig:v2_200} (Color online) $v_{2}(p_{\rm T})$ results at $\sqrt{s_{\rm NN}} = $ 200 GeV from 0\%--80\% central Au+Au collisions for identified particles ($\pi^{\pm}, K^{\pm}, p, \bar{p}$ and $\phi$) in comparison to RHIC data. The open symbols in the panels of $\pi^{\pm},K^{\pm},p$ and $\bar{p}$ are from PHENIX~\cite{Adler:2003kt} and the panel of $\phi$ are from STAR~\cite{Abelev:2007rw} and the lines are from AMPT calculation.}
\end{figure}
%-----------------------------------------------------------------------------------------------

%-----------------------------------------------------------------------------------------------
% figure 5
\begin{figure}[h!]
  \centering
  \includegraphics[width=0.4\textwidth]{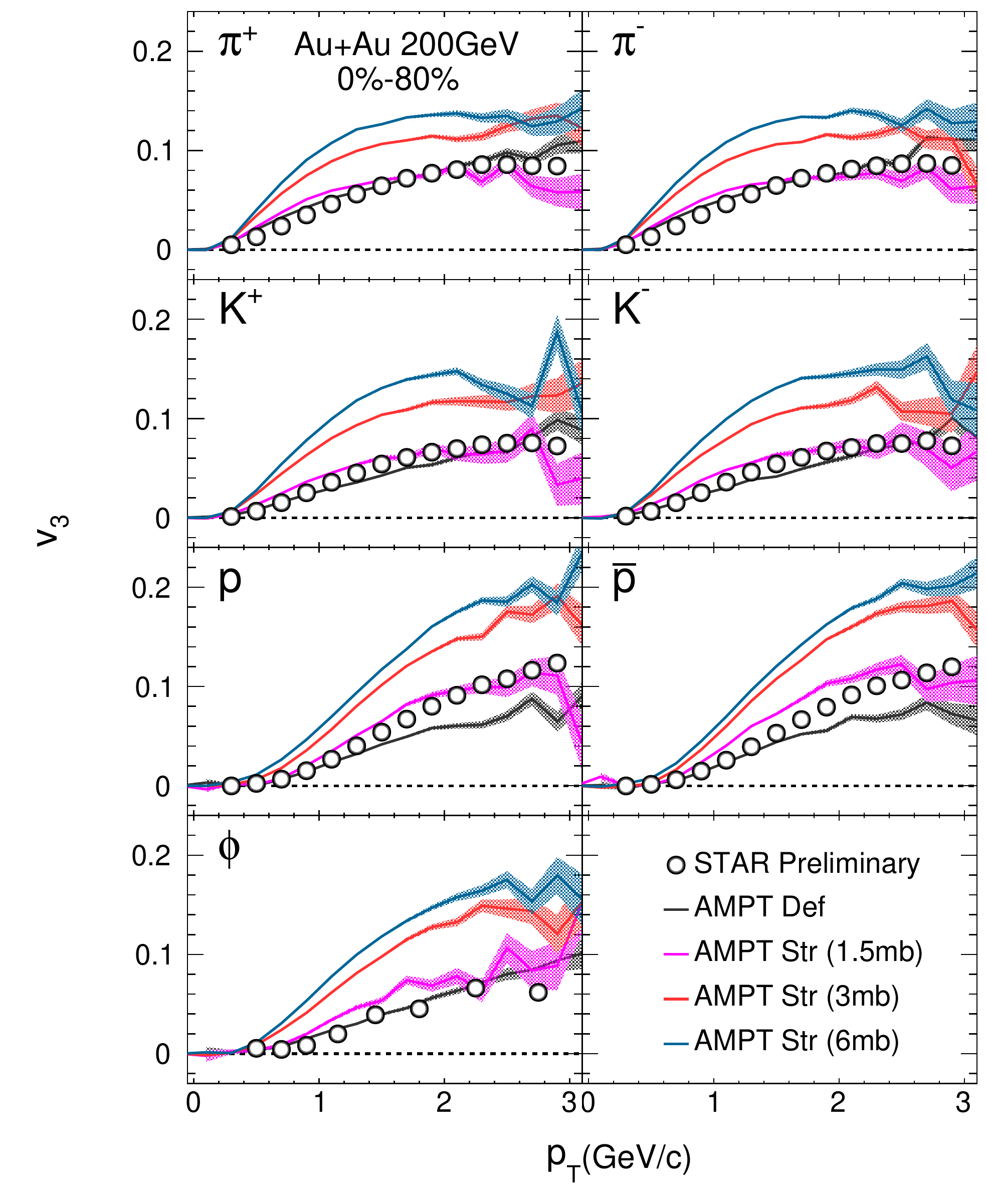}%
  \caption{\label{fig:v3_200} (Color online) $v_{3}(p_{\rm T})$ results at $\sqrt{s_{\rm NN}} = $ 200 GeV from 0\%--80\% central Au+Au collisions for identified particles ($\pi^{\pm}, K^{\pm}, p, \bar{p}$ and $\phi$) in comparison to STAR data.}
\end{figure}
%-----------------------------------------------------------------------------------------------

\section{\label{sec:result} Results}
Identical analysis steps are used for simulation and data collected from the experiments~\cite{Adamczyk:2013gw,Sun:2014yqi}. Finial state charged particles with 0.2 $ < p_{\rm T} < $ 2.0 GeV/c and $\left|\eta\right| < $ 1.0, are used to reconstruct the event plane. The $\eta$ range of the two $\eta-$sub events are -1.0 $ < \eta < $ -0.05 and 0.05 $ < \eta < $ 1.0, respectively. The elliptic and triangular flow are calculated within -1.0 $ < \eta < $ 1.0, and the auto-correlation is naturally avoided by the $\eta$-sub event plane method. For all the plots in this section, only statistical errors are shown.

\subsection{\label{sub:ampt_flow} Elliptic Flow and Triangular Flow at RHIC}
Figure~\ref{fig:res} presents the second and third harmonic event plane resolutions from the AMPT string melting model for parton scattering cross sections of 1.5 mb, 3 mb and 6 mb from 0\%-80\% central Au+Au collisions at $\sqrt{s_{\rm NN}} = $ 11.5, 39 and 200 GeV by using the $\eta$-sub event plane method. The event plane resolution shows a clear parton scattering cross section and energy dependence: both second and third harmonic event plane resolution increases with increasing parton scattering cross section and energy. 

The third harmonic event plane resolution is about two times smaller and peaks more central than the second harmonic event plane resolution, which is comparable to the one from experiment. For $\sqrt{s_{\rm NN}} = $ 11.5 and 39 GeV, $v_{3}$ can only be calculated up to 0\%--60\% (0\%--70\%) for some of the parton scattering cross sections, since we only have event plane resolution up to these centrality.

Figure~\ref{fig:v2_11}, Fig.~\ref{fig:v2_39} and Fig.~\ref{fig:v2_200} show the comparison between data from experiment and from the AMPT model for $v_{2}(p_{\rm T})$ of identified particles ($\pi^{\pm}, K^{\pm}, p, \bar{p}$ and $\phi$) at $\sqrt{s_{\rm NN}} = $ 11.5, 39 and 200 GeV from 0\%--80\% central Au+Au collisions.
The data from experiment are taken from Ref.~\cite{Adamczyk:2013gw,Adler:2003kt}. For all three energies it is observed that the elliptic flow increases with increasing parton scattering cross section. The elliptic flow from default AMPT is lower than the corresponding results from experiment, which indicates that the parton scattering process is important to produce a large elliptic flow as observed in experiment data.
Previous studies show that a 6-10 mb parton scattering cross section is needed to reproduce the elliptic flow observed in experiment data~\cite{Lin:2001zk,Chen:2004dv}. In those studies the elliptic flow was calculated relative to the reaction plane. 

By using the  $\eta$-sub event plane method most of the experiment data points at low transverse momenta are between AMPT default and AMPT string melting with a 3 mb parton scattering cross section. At intermediate $p_{\rm T}$ some particle species, e.g. $\pi^{\pm}$, are systematically above AMPT string melting with a 3 mb parton scattering cross section. That means previous studies have underestimated the elliptic flow by using the reaction plane instead of the event plane. Latter one fluctuates event-by-event around the reaction plane. The collective flow measured in experiment are not mean values, but closer to a root-mean-sqaure~\cite{Ollitrault:2009ie}, therefore, the  event-by-event fluctuations did not cancel each other but give a positive contribution to the measured collective flow, which makes the collective flow relative to the event plane is always larger compare to the collective flow relative to the reaction plane.

Figure~\ref{fig:v3_200} shows the comparison between experiment data~\cite{Sun:2014yqi} and AMPT model calculations for $v_{3}(p_{\rm T})$  at $\sqrt{s_{\rm NN}} = $ 200 GeV from 0\%--80\% central Au+Au collisions of identified particles ($\pi^{\pm},K^{\pm},p,\bar{p}$ and $\phi$). A similar parton scattering cross section dependence is observed as for the elliptic flow, but here the dependence is stronger: $v_{3}$ is about 40\% lower than $v_{2}$ with 1.5 mb parton scattering cross section but comparable with $v_{2}$ with a 6 mb parton scattering cross section. The driving force behind $v_{2}$ is the almond shape of initial nuclei overlap region, therefore, $v_{2}$ has a relative large value compare to $v_{3}$ which is mainly generated by initial state fluctuations~\cite{Alver:2010gr}. This effect can be observed by comparing the $v_{2}$ and $v_{3}$ values from AMPT default model calculations in Fig.~\ref{fig:v2_200} and Fig.~\ref{fig:v3_200}. The higher parton scattering cross section makes the transportation from the initial coordinate space to the final momentum space more efficient, which increases the values for both $v_{2}$ and $v_{3}$. The relative increase of $v_{3}$  is larger than for $v_{2}$, since the effects of the parton scattering is more important for $v_{3}$.
The triangular and elliptic flow from experiment can be well reproduced by AMPT default and AMPT string melting  with a 1.5 mb parton scattering cross section, see Fig.~\ref{fig:v2_200} and Fig.~\ref{fig:v3_200}.
It is worth to be noticed here that we did not find out a single parton scattering cross section to describe all particle species.
This is because the collective flow results from AMPT model depend on both the magnitude and the distribution of parton scattering cross section.
Recent analysis~\cite{Xu:2011fe} showed that the experiment elliptic flow results can be described by the string melting AMPT model with a smaller but more isotropic parton scattering cross section.
Thus, by carefully tuning the parton scattering cross section and its distribution, we might achieve a roughly description of all parton species, but this is beyond the goal of this paper.

\subsection{\label{sub:ampt_diff} Differential $v_{2}(p_{\rm T})/v_{3}(p_{\rm T})$ Ratio from AMPT Model}

\begin{figure}[h]
  \centering
  \includegraphics[width=0.45\textwidth]{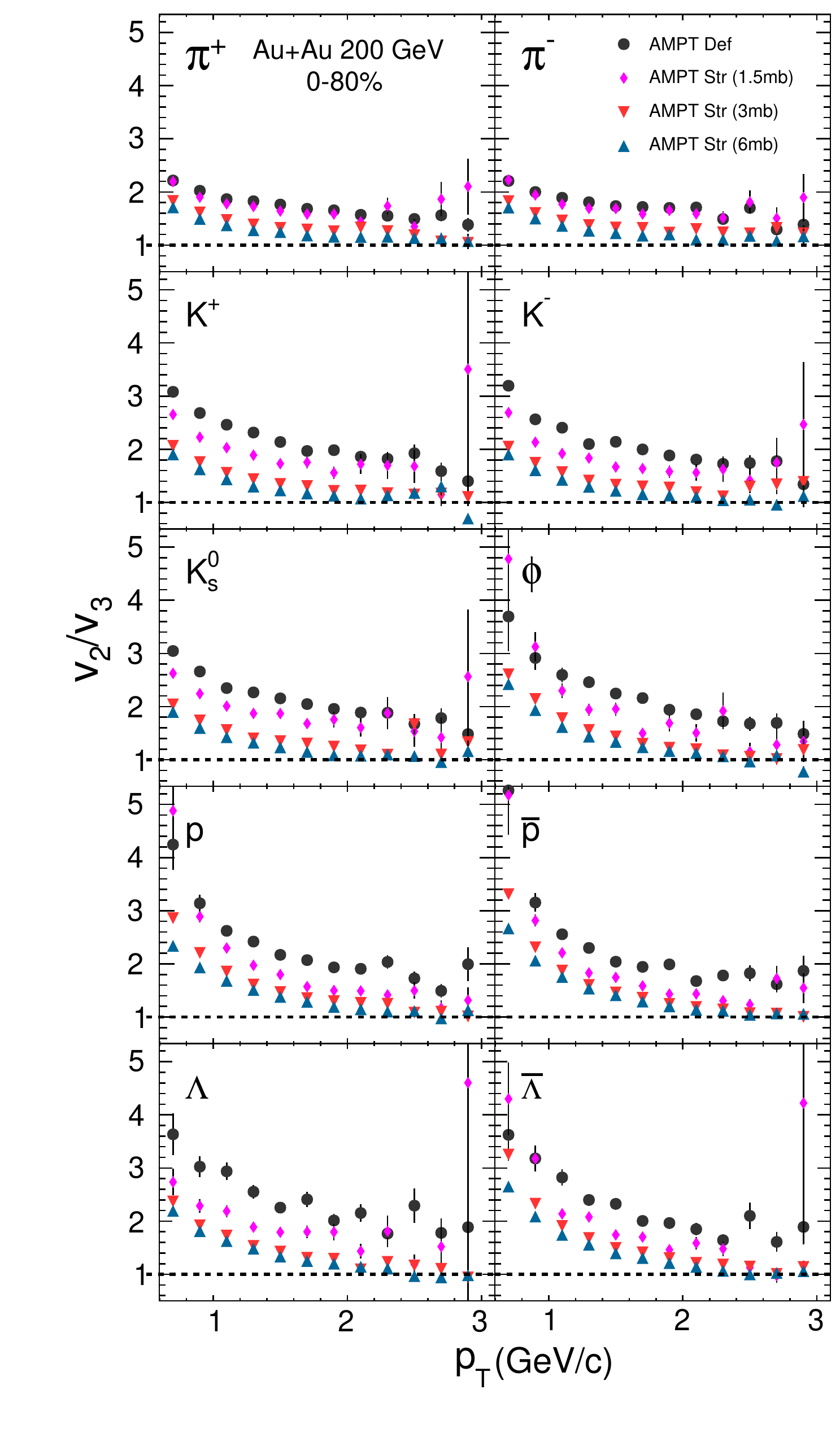}%
  \caption{\label{fig:Diff} (Color online) AMPT calculation of differential $v_{2}(p_{\rm T})/v_{3}(p_{\rm T})$ ratios for identified particles ($\pi^{\pm}$, $K^{\pm}$, $K^{0}_{s}$, $\phi$, $p$, $\bar{p}$, $\Lambda$ and $\bar{\Lambda}$) from 0\%--80\% central Au+Au collisions at $\sqrt{s_{\rm NN}} = $ 200 GeV.}
\end{figure}

Figure~\ref{fig:Diff} depicts AMPT model calculations for the transverse momentum dependent differential ratio $v_{2}(p_{\rm T})/v_{3}(p_{\rm T})$ for identified particles ($\pi^{\pm}, K^{\pm}, K^{0}_{s}, \phi, p, \bar{p}, \Lambda$ and $\bar{\Lambda}$) from 0\%--80\% central Au+Au collisions at $\sqrt{s_{\rm NN}} = $ 200 GeV.
The differential ratio decreases significantly at low transverse momenta ($p_{\rm T} < $ 1.5 GeV/c) and becomes flat at intermediate transverse momenta (1.5 $< p_{\rm T} <$ 3.0 GeV/c).
The $p_{\rm T}$ dependent differential ratio shows an anti-correlation with parton scattering cross section for all particle species: the higher the parton scattering cross section, the lower the $v_{2}(p_{\rm T})/v_{3}(p_{\rm T})$ ratio.
This is in agreement with a larger increase of $v_{3}$ with increasing parton scattering cross section compare to $v_{2}$ discussed in Fig.~\ref{fig:v2_200} and Fig.~\ref{fig:v3_200}.
This effect can also be understood in a hydrodynamic frame work. The higher parton scattering cross section in the AMPT string melting model is equivalent to a lower viscosity in viscous hydrodynamics~\cite{Gyulassy:1997ib,Zhang:1999rs}. From this study it is known that $v_{3}$ is more sensitive to  viscosity than $v_{2}$~\cite{Alver:2010dn}, therefore a lower viscosity (higher parton scattering cross section) leads to a lower $v_{2}(p_{\rm T})/v_{3}(p_{\rm T})$ ratio.

\subsection{\label{sub:ampt_inte} Integrated $v_{2}/v_{3}$ Ratio from AMPT Model}

\begin{figure}[h]
  \centering
  \includegraphics[width=0.55\textwidth]{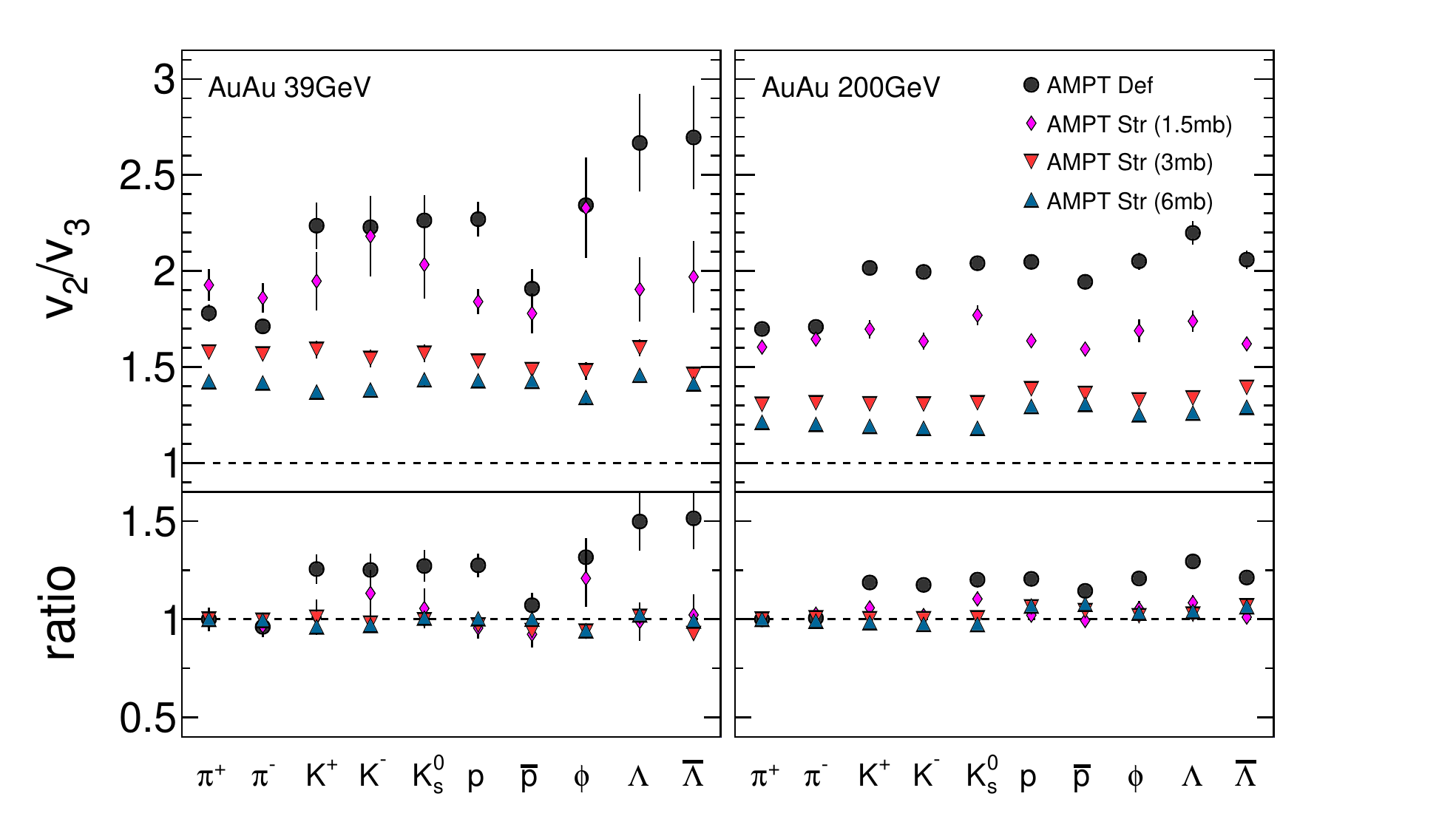}%
  \caption{\label{fig:Inte} (Color online) Upper panel: AMPT calculation of the integrated $v_{2}/v_{3}$ ratio for identified particles ($\pi^{\pm}, K^{\pm}, K^{0}_{s}, \phi, p, \bar{p}, \Lambda$ and $\bar{\Lambda}$) from 0\%--80\% central Au+Au collisions at $\sqrt{s_{\rm NN}} = $ 39 and 200 GeV. Bottom panel: the ratio between integrated $v_{2}/v_{3}$ of all particles species to that of $\pi^{+}$}
\end{figure}

Figure~\ref{fig:Inte} shows the AMPT calculation for the integrated $v_{2}/v_{3}$ ratio for identified particles from 0\%--80\% central Au+Au collisions at $\sqrt{s_{\rm NN}} = $ 39 and 200 GeV. The integration region is limited to 1.5 $< p_{\rm T} <$ 2.8 GeV/$c$ to avoid the steep decrease in the differential ratio and the large error region.
Please note, the triangular flow used to calculate the $v_{2}(p_{\rm T})/v_{3}(p_{\rm T})$ ratio for 1.5 mb and 3 mb parton scattering cross section are taken from 0\%--60\% and 0\%--70\% due to the event plane resolution limits, as shown in Fig.~\ref{fig:res}. The other of $v_{3}$ values are taken from 0\%--80\% results.

A similar anti-correlation between the integrated $v_{2}/v_{3}$ ratio and the parton scattering cross section is observed as in the differential $v_{2}(p_{\rm T})/v_{3}(p_{\rm T})$ ratio: the larger the parton scattering cross section, the smaller the integrated $v_{2}/v_{3}$ ratio. In addition, for the same parton scattering cross section, an energy dependence is observed: the lower the energy the larger the integrated $v_{2}/v_{3}$ ratio.
We further observe that the integrated $v_{2}/v_{3}$ ratio calculated by string melting AMPT shows no mass (particle type) dependence, the ratio between integrated $v_{2}/v_{3}$ of all particle species to that of $\pi^{+}$ are consistent with 1 within 10\%.
On the other hand the ratios of $\pi^{\pm}$ from default AMPT are lower than the rest of the particle species at both $\sqrt{s_{\rm NN}} = $ 39 and 200 GeV.
This indicates that the quark coalescence mechanism might play an important role in flattening the $v_{2}/v_{3}$ ratio.
The no mass (particle type) dependence of the $v_{2}/v_{3}$ ratio was predicted by viscous hydrodynamics, for fast particles (particles with large $p_{\rm T}$). The $v_{2}/v_{3}$ ratio is constant and does not depend on the mass of the particle~\cite{Lang:2013oba}. 
Based on the viscous hydrodynamics in Ref.~\cite{Lang:2013oba}, the collective flow depends on dissipative contributions and the Fourier coefficient ($V_{n}$) of the flow velocity distribution relative to the event plane, the dissipative contributions cancelled each other in $v_{2}/v_{3}$ ratio calculation and only $V_{2}/V_{3}$ left which does not depend on the mass of different particle species.
Since the parton scattering cross section is related to the viscosity in hydrodynamics, the magnitude of the $v_{2}/v_{3}$ ratio might be usable to quantify the viscosity of the system.

\section{\label{sec:sum} Summary}
The AMPT calculations of elliptic flow at $\sqrt{s_{\rm NN}} = $ 11.5, 39 and 200 GeV and triangular flow at $\sqrt{s_{\rm NN}} = $ 200 GeV of identified particles ($\pi^{\pm}, K^{\pm}, p, \bar{p}$ and $\phi$) from 0\%--80\% central Au+Au collisions were presented. Most of results from experiments can be reproduced with a parton scattering cross section between 1.5 and 3 mb by following the same analysis method as used in the experiments.
An anti-correlation between the parton scattering cross section for both the differential and integrated $v_{2}/v_{3}$ ratio is observed. This is due to a different sensitivity of $v_{2}$ and $v_{3}$ to the parton scattering cross section.
A mass (particle type) independence of the integrated $v_{2}/v_{3}$ ratio is observed from string melting AMPT, but not for the default AMPT. 
The no mass (particle type) dependence of the $v_{2}/v_{3}$ ratio also can be understood in the viscous hydrodynamic framework. Latter one suggests that the magnitude of the ratio might be used to quantify the viscosity of the system.

%\appendix
%\section{}

\begin{acknowledgments}
 We thank Dr. Nu Xu for his great idea and important discussions and Dr. Guoliang Ma for his help on AMPT model calculations.
 This work was supported by the National Natural Science Foundation of China under grant No.11475070 and U1332125 and the Program for Innovation Research of Science in Harbin Institute of Technology (PIRS OF HIT B201408).
\end{acknowledgments}

\bibliography{AMPT_ratio}

\end{document}